\documentclass[conference,letterpaper]{IEEEtran}

\usepackage[utf8]{inputenc}
\usepackage[T1]{fontenc}
\usepackage{url}
\usepackage{ifthen}
\usepackage{cite}
\usepackage[cmex10]{amsmath}

\usepackage{bm}
\usepackage{amsfonts}
\usepackage{amsthm}
\usepackage{graphicx,ifthen}
\usepackage{mathtools}
\usepackage{multirow}
\usepackage{lipsum}

\usepackage{array}
\usepackage{fixltx2e}
\usepackage{color}
\usepackage{amssymb}
\usepackage{t1enc}
\usepackage{alltt}
\usepackage{epsfig}
\usepackage{mathrsfs}
\usepackage[scaled]{helvet}
\usepackage{booktabs}
\usepackage{wasysym}
\usepackage{arydshln}
\usepackage{enumerate}
\usepackage{bbm}

\allowdisplaybreaks[3]
\newcommand{\e}{{\rm e}}

\newcommand{\dd}{\,\mathrm{d}}
\newtheorem{lemma}{Lemma}
\newtheorem{proposition}{Proposition}
\newcommand{\ra}[1]{\renewcommand{\arraystretch}{#1}}
\makeatletter
\def\hrulefill{\leavevmode\leaders\hrule height 1.0pt\hfill\kern\z@}

\newcounter{tempEqNo}

\begin{document}
\title{Ergodic MIMO Mutual Information:\\Twenty Years After Emre Telatar}

\author{%
\IEEEauthorblockN{Lu Wei}
\IEEEauthorblockA{Department of Electrical and Computer Engineering\\
                     University of Michigan - Dearborn\\
                     Dearborn, MI 48128, USA\\
                     Email: luwe@umich.edu}}

\maketitle

\begin{abstract}
In the celebrated work of Emre Telatar in the year 1999 (14274 citations to date), it was shown that the expected value of the mutual information
\begin{equation*}
\mathrm{I}=\ln\det\left(\mathbf{I}_{m}+\frac{1}{t}\mathbf{HH}^{\dag}\right)
\end{equation*}
of an $m\times n$ MIMO Rayleigh channel matrix $\mathbf{H}$ with a SNR $1/t$ can be represented as an integral involving Laguerre polynomials. We show, in this work, that Telatar's integral representation can be explicitly evaluated to a finite sum of the form
\begin{equation*}
\mathbb{E}\!\left[\mathrm{I}\right]=\sum_{k=0}^{n+m-3}a_{k}t^{k}+\e^{t}~\text{\normalfont Ei}(-t)\sum_{k=0}^{n+m-2}b_{k}t^{k},
\end{equation*}
where $\text{\normalfont Ei}(-t)$ is the exponential integral and $a_{k}$, $b_{k}$ are known constants that do not dependent on $t$. The renewed interest in this classical information theory problem came from, quite surprisingly, the recent development in quantum information theory.
\end{abstract}

\section{Introduction}
The notion of mutual information is among the most important quantities in information theory. Its supremum value provides the fundamental performance measure, the channel capacity. For Multiple-Input-Multiple-Output (MIMO) Rayleigh fading channel, that corresponds to a rich scattering environment, it is a well-known result~\cite[Theorem 1]{1999Telatar} that the ergodic capacity is achieved when the input follows a complex Gaussian distribution with zero mean and covariance proportional to an identity matrix. The operational meaning of ergodic capacity is that the transmission of each codeword has experienced sufficient many channel realizations, i.e., a fast fading scenario. Mathematically, the ergodic capacity of Rayleigh fading channels is given by the expected value of the mutual information as~\cite{1999Telatar}
\begin{equation}\label{eq:I}
\mathrm{I}=\ln\det\left(\mathbf{I}_{m}+\frac{1}{t}\mathbf{HH}^{\dag}\right),
\end{equation}
where $1/t$ is the Signal-to-Noise Ratio (SNR) and the entries of the $m\times n$ channel matrix $\mathbf{H}$ spanned by the pathes between $n$ transmit and $m$ receive antennas are independent and follow a complex Gaussian distribution with zero mean and unit variance. Here, $\ln(\cdot)$ is the natural logarithm with the unit of~(\ref{eq:I}) being nats/second/Hz, $\det(\cdot)$ is the matrix determinant, and it can be assumed without loss of generality that $m\leq n$.

Denote the eigenvalues of the Hermitian matrix $\mathbf{HH}^{\dag}$ by $0<\lambda_{m}<\dots<\lambda_{2}<\lambda_{1}<\infty$, the mutual information is rewritten as
\begin{equation}\label{eq:II}
\mathrm{I}=\sum_{i=1}^{m}\ln\left(1+\frac{1}{t}\lambda_{i}\right).
\end{equation}
By the symmetry of~(\ref{eq:II}) in all the eigenvalues, the ergodic mutual information is evaluated as
\begin{equation}\label{eq:E1p}
\mathbb{E}\!\left[\mathrm{I}\right]=m\mathbb{E}\!\left[\ln\left(1+\frac{1}{t}\lambda\right)\right],
\end{equation}
where the right-hand-side expectation is now taken over the density of an arbitrary eigenvalue (a.k.a. the one-point density) given by~\cite{Mehta,1999Telatar}
\begin{equation}\label{eq:1p}
p(\lambda)=\frac{1}{m}\e^{-\lambda}\lambda^{n-m}\sum_{k=0}^{m-1}\frac{k!}{(n-m+k)!}\left(L_{k}^{(n-m)}(\lambda)\right)^{2}
\end{equation}
with $\lambda\in[0,\infty)$. Here, $L_{k}^{(\alpha)}(x)$ is the Laguerre orthogonal polynomial of degree $k$ in $x$ that satisfies the orthogonality relation~\cite{Mehta}
\begin{equation}\label{eq:oc}
\int_{0}^{\infty}\!\!x^{\alpha}\e^{-x}L_{k}^{(\alpha)}(x)L_{l}^{(\alpha)}(x)\dd{x}=\frac{(\alpha+k)!}{k!}\delta_{kl},
\end{equation}
where $\delta_{kl}$ is the Kronecker delta function. Inserting~(\ref{eq:1p}) into~(\ref{eq:E1p}), one arrives at an integral representation of the ergodic mutual information of MIMO Rayleigh channels as
\begin{eqnarray}\label{eq:T99}
\mathbb{E}\!\left[\mathrm{I}\right]=\sum_{k=0}^{m-1}\frac{k!}{(n-m+k)!}\int_{0}^{\infty}\ln\left(1+\frac{1}{t}\lambda\right)\times\nonumber\\
\e^{-\lambda}\lambda^{n-m} \left(L_{k}^{(n-m)}(\lambda)\right)^{2}\dd\lambda.
\end{eqnarray}
In fact, the above result was stated as Theorem 2 in the seminal work by Telatar~\cite{1999Telatar}.

In this paper, we show that Telatar's representation~(\ref{eq:T99}) can be calculated to an explicit form involving, up to a factor $\e^{t}~\text{\normalfont Ei}(-t)$, finite polynomials in $t$ as presented in the next section. The corresponding proof will be outlined in Section~\ref{sec:proof}. In Section~\ref{sec:con}, we point out some further work that may be carried out based on the obtained results.

\begin{figure*}[!t]
\vspace*{20pt}
\setcounter{tempEqNo}{\value{equation}}
\setcounter{equation}{8}
\begin{align}
a_{k}&=\begin{dcases}\label{eq:a}
\sum_{j=0}^{m-1}\sum_{i=j}^{2m-2}(i+n-m)!~c_{ij}\sum_{s=1}^{i+n-m}\frac{1}{s}, & \quad k=0\\
\frac{(-1)^{k}}{kk!}\sum_{j=0}^{m-1}\sum_{i=k-n+m+1}^{2m-2}\big((i+n-m)!-k!(i+n-m-k)!\big)~c_{ij}, & \quad k=1,\dots,n+m-3
\end{dcases}\\[14pt]
b_{k}&=\begin{dcases}\label{eq:b}
-\frac{(-1)^{k}m}{k!}, & \quad k=0,\dots,n-m\\
-\frac{(-1)^{k}}{k!}\sum_{j=0}^{m-1}\sum_{i=k-n+m}^{2m-2}(i+n-m)!~c_{ij}, & \quad k=n-m+1,\dots,n+m-2
\end{dcases}\\[14pt]
c_{ij}&=\frac{n!m!(-1)^{i}(2nj+j-ni+n-m+1)}{j!(i-j)!(n-m+1+j)!(n-m+1+i-j)!(m-1-j)!(m-i+j)!}\label{eq:c}\\[-11pt]\nonumber
\end{align}
\setcounter{equation}{\value{tempEqNo}}
\hrulefill\vspace*{14pt}
\end{figure*}

\section{Main Result}
The main result of this paper is summarized in the following proposition.
\begin{proposition}\label{p:1}
The ergodic mutual information of MIMO Rayleigh fading channels with $n$ transmit antennas, $m$ receive antennas, and any SNR value $1/t$, i.e., the expected value of the random variable~(\ref{eq:I}), is given by
\begin{equation}\label{eq:main}
\mathbb{E}\!\left[\mathrm{I}\right]=\sum_{k=0}^{n+m-3}a_{k}t^{k}+\e^{t}~\text{\normalfont Ei}(-t)\sum_{k=0}^{n+m-2}b_{k}t^{k},
\end{equation}
where
\begin{equation}\label{eq:Ei}
\text{\normalfont Ei}(x)=-\int_{-x}^{\infty}\frac{\e^{-s}}{s}\dd s,
\end{equation}
is the exponential integral function. The constants $a_{k}$ and $b_{k}$ are given by~(\ref{eq:a}) and~(\ref{eq:b}), respectively, as shown on top of this page.
\end{proposition}

Proposition~\ref{p:1} reveals that the ergodic mutual information of MIMO Rayleigh channels consists of finite polynomials in the inverse of SNR $t$ with a common factor $\e^{t}~\text{\normalfont Ei}(-t)$ independent of the channel dimensions. The degree of the polynomials depends essentially on the sum of the channel dimensions $m+n$. The coefficients of the polynomials $a_{k}$ and $b_{k}$ in turn are given by finite sums of factorial functions. Clearly, these coefficients can be efficiently computed for any dimensions $m$ and $n$ with no special functions involved. The unveiled simple structure~(\ref{eq:main}) of the ergodic mutual information is far from clear a priori. The functional dependence on the SNR is explicitly displayed. This facilitates further analysis on the statistical behavior of mutual information when the SNR is modeled as a random variable. The SNR variations may result from, for example, transceiver schemes or physical channels.

Before proving the Proposition~\ref{p:1}, we list in Table~\ref{t:eg} some examples computed from the derived formula~(\ref{eq:main}) of different channel dimensions as shown on top of the next page. We also point out that extensive simulation has been performed that numerically verified the main result~(\ref{eq:main}). Interested readers may contact the author to obtain the code implemented in Mathematica.

\begin{table*}[!t]
\caption{Ergodic Mutual Information~(\ref{eq:main}): Some Examples}\centering
\ra{2.6}
\begin{tabular}{ccc}
\toprule
\\[-26pt]\multirow{3}{*}{$m=2$} & $n=2$ & $1-t-\e^{t}~\text{\normalfont Ei}(-t)\left(2+t^{2}\right)$ \\
                       & $n=4$ & $\frac{1}{6}\Big(20-6t-t^{2}-t^{3}-\e^{t}~\text{\normalfont Ei}(-t)\left(12-12t+6t^{2}+2t^{3}+t^{4}\right)\!\Big)$ \\
                       & $n=6$ & $\frac{1}{120}\Big(524-180t+48t^{2}-8t^{3}-3t^{4}-t^{5}-\e^{t}~\text{\normalfont Ei}(-t)\left(240-240t+120t^{2}-40t^{3}+10t^{4}+4t^{5}+t^{6}\right)\!\Big)$ \\ \hline
\multirow{3}{*}{$m=4$} & $n=4$ & $\frac{1}{36}\Big(156-156t-96t^{2}-56t^{3}-11t^{4}-t^{5}-\e^{t}~\text{\normalfont Ei}(-t)\left(144+216t^{2}+144t^{3}+66t^{4}+12t^{5}+t^{6}\right)\!\Big)$ \\
                       & \multirow{2}{*}{$n=6$} & $\frac{1}{720}\Big(5544-1440t-720t^{2}-1600t^{3}-756t^{4}-186t^{5}-21t^{6}-t^{7}-\e^{t}~\text{\normalfont Ei}(-t)\times$ \\
                       & & $\left(2880-2880t+1440t^{2}+1920t^{3}+2200t^{4}+924t^{5}+206t^{6}+22t^{7}+t^{8}\right)\!\Big)$ \\
\bottomrule
\end{tabular}\label{t:eg}
\end{table*}

\section{Proof of the Main Result}\label{sec:proof}
The proof of Proposition~\ref{p:1} involves two steps with the starting point being Telatar's representation~(\ref{eq:T99}). The first step is to manipulate the integrand of~(\ref{eq:T99}) to a form such that the integral can be conveniently calculated. The second step is to simplify, with the help of special function theory, the multiple nested summations produced in the first step. This two-step procedure has been recently utilized in solving a related problem\footnote{Instead of mutual information, the random variable of interest in~\cite{Wei2017} is $\sum_{i=1}^{m}\lambda_{i}\ln\lambda_{i}$ known as the von Neumann entropy. Here, the presence of $\lambda_{i}$ in front of the logarithm is not essential, which is absorbed in the density~(\ref{eq:1p}). Thus, the considered mutual information~(\ref{eq:II}) is a more generalized quantity, i.e., a one parameter $t$ deformation to the von Neumann entropy, cf.~(\ref{eq:III}).} in quantum information theory~\cite{Wei2017}.

For the first step, we start by rewriting~(\ref{eq:II}) as
\addtocounter{equation}{3}
\begin{equation}\label{eq:III}
\mathrm{I}=-m\ln t+\sum_{i=1}^{m}\ln\left(t+\lambda_{i}\right)
\end{equation}
so that by~(\ref{eq:E1p}) the expected value becomes
\begin{equation}\label{eq:En1p}
\mathbb{E}\!\left[\mathrm{I}\right]=-m\ln t+m\int_{0}^{\infty}\ln\left(t+\lambda\right)p(\lambda)\dd t.
\end{equation}
To proceed, we use the fact that the one-point density~(\ref{eq:1p}) utilized in Telatar's result~(\ref{eq:T99}) admits a more convenient form~\cite{Mehta} with no summation involved
\begin{eqnarray}\label{eq:n1p}
p(\lambda)&=&\frac{(m-1)!}{(n-1)!}\lambda^{n-m}\e^{-\lambda}\bigg(\!\left(L_{m-1}^{(n-m+1)}(\lambda)\right)^{2}-\nonumber\\
&&L_{m-2}^{(n-m+1)}(\lambda)L_{m}^{(n-m+1)}(\lambda)\bigg).
\end{eqnarray}
Inserting~(\ref{eq:n1p}) into~(\ref{eq:En1p}) and by changing the variable $x\to x-t$, we have
\begin{equation}\label{eq:EIi}
\mathbb{E}\!\left[\mathrm{I}\right]=-m\ln t+\frac{m!}{(n-1)!}\left(A_{m-1,m-1}\left(t\right)-A_{m-2,m}\left(t\right)\right),
\end{equation}
where we denote
\begin{eqnarray}\label{eq:I12}
A_{s,t}\left(t\right)&=&e^{t}\int_{t}^{\infty}L_{s}^{(n-m+1)}(x-t)L_{t}^{(n-m+1)}(x-t)\times \nonumber\\
&&(x-t)^{n-m}\e^{-x}\ln x\dd x.
\end{eqnarray}
Since the product $(x-t)^{n-m}L_{s}^{(n-m+1)}(x-t)L_{t}^{(n-m+1)}(x-t)$ is a polynomial in $x$, to evaluate~(\ref{eq:I12}) it remains to see if an integral of the form $\int_{t}^{\infty}x^{k}\e^{-x}\ln x\dd x$ can be calculated. This turns out to be possible, and the result is given in the lemma below.
\begin{lemma}\label{l:1}
For any non-negative integer $k$, we have
\begin{equation}\label{eq:L1}
\int_{t}^{\infty}x^{k}\e^{-x}\ln x\dd x=\Gamma\left(k+1,t\right)\ln t+k!\sum_{s=0}^{k}\frac{\Gamma\left(s,t\right)}{s!},
\end{equation}
where
\begin{equation}\label{eq:iG}
\Gamma\left(s,t\right)=(s-1)!~\e^{-t}\sum_{i=0}^{s-1}\frac{t^{i}}{i!}
\end{equation}
defines the incomplete Gamma function for a positive integer~$s$.
\end{lemma}
The proof of Lemma~\ref{l:1} is in the Appendix. By Lemma~\ref{l:1} and the series form of the Laguerre polynomial
\begin{equation}\label{eq:L}
L_{k}^{(\alpha)}(x)=\sum_{i=0}^{k}(-1)^{i}\binom{\alpha+k}{k-i}\frac{x^i}{i!},
\end{equation}
the integral~(\ref{eq:I12}) is computed as in~(\ref{eq:Ast}) shown on top of the next page.
\begin{figure*}[!t]
\vspace*{20pt}
\begin{eqnarray}
A_{s,t}\left(t\right)&=&\e^{t}\int_{t}^{\infty}\sum_{i=0}^{s+t}\sum_{j=0}^{i}\frac{(-1)^i}{j!(i-j)!}\binom{n-m+1+t}{t-j}\binom{n-m+1+s}{s-i+j}\sum_{k=0}^{i+n-m}\binom{i+n-m}{k}(-t)^{i+n-m-k}x^{k}\e^{-x}\ln x\dd x \nonumber\\
&=&\e^{t}\sum_{i=0}^{s+t}\sum_{j=0}^{i}\frac{(-1)^i}{j!(i-j)!}\binom{n-m+1+t}{t-j}\binom{n-m+1+s}{s-i+j}\sum_{k=0}^{i+n-m}\binom{i+n-m}{k}(-t)^{i+n-m-k}\times \nonumber \\
&&\left(\Gamma\left(k+1,t\right)\ln t+k!\sum_{s=0}^{k}\frac{\Gamma\left(s,t\right)}{s!}\right) \label{eq:Ast}
\end{eqnarray}
\hrulefill\vspace*{9pt}
\end{figure*}
Inserting~(\ref{eq:Ast}) into~(\ref{eq:EIi}) and by factoring out from $A_{m-1,m-1}$ and $A_{m-2,m}$ the common term $c_{ij}$ as shown in~(\ref{eq:c}), we obtain
\begin{eqnarray}
\mathbb{E}\!\left[\mathrm{I}\right]\!\!\!\!&=\!\!\!\!&-m\ln~\!\!t+\e^{t}\sum_{i=0}^{2m-2}\sum_{j=0}^{i}c_{ij}\sum_{k=0}^{i+n-m}\binom{i+n-m}{k}\times \nonumber \\
&&\mkern-25mu (-t)^{i+n-m-k}\left(\Gamma(k+1,t)\ln~\!\!t+k!\sum_{s=0}^{k}\frac{\Gamma\left(s,t\right)}{s!}\right). \label{eq:EIi2}
\end{eqnarray}

We have so far carried out the first step of the proof, i.e., computing the integral in Telatar's formula~(\ref{eq:T99}). The second step is to simplify the obtained expression~(\ref{eq:EIi2}). We first show that the two terms involving $\ln t$ in~(\ref{eq:EIi2}) cancel, i.e., the equality below holds
\begin{eqnarray}
m&=&\e^{t}\sum_{i=0}^{2m-2}\sum_{j=0}^{i}c_{ij}\sum_{k=0}^{i+n-m}\binom{i+n-m}{k}\times \nonumber \\
&&(-t)^{i+n-m-k}\Gamma(k+1,t). \label{eq:lnt}
\end{eqnarray}
The above equality is established by sequentially applying the summation formulas below
\begin{eqnarray}
\sum_{k=0}^{N}\binom{N}{k}(-t)^{-k}\Gamma(k+1,t)&=&\left(-\frac{1}{t}\right)^{N}N!\e^{-t},\\
\sum_{k=0}^{N}\frac{(-N)_{k}(a)_{k}}{k!(b)_{k}}&=&\frac{(b-a)_{N}}{(b)_{N}}, \label{eq:CV}
\end{eqnarray}
where the first formula follows from the definition~(\ref{eq:iG}) and the second one is known as the Chu-Vandermonde identity (see, e.g.,~\cite{Wei2017}) with $(a)_{n}=\Gamma(a+n)/\Gamma(a)$ denoting the Pochhammer's symbol. With the result~(\ref{eq:lnt}) and the fact that $\Gamma(0,t)=-\text{\normalfont Ei}(-t)$ (cf.~(\ref{eq:Ei}) and~(\ref{eq:iGi})), we can write~(\ref{eq:EIi2}) as
\begin{eqnarray}
\mathbb{E}\!\left[\mathrm{I}\right]\!\!\!\!&=\!\!\!\!&\e^{t}\sum_{i=0}^{2m-2}\sum_{j=0}^{i}c_{ij}\sum_{k=0}^{i+n-m}\binom{i+n-m}{k}\times \nonumber \\
&&\mkern-25mu (-t)^{i+n-m-k}k!\left(\sum_{s=1}^{k}\frac{\Gamma\left(s,t\right)}{s!}-\text{\normalfont Ei}(-t)\right), \label{eq:EIi3}
\end{eqnarray}
where the coefficients $a_{k}$ and $b_{k}$ are now related to the coefficients of the first and second term in the last bracket in~(\ref{eq:EIi3}), respectively. Putting aside the factor $\e^{t}\text{\normalfont Ei}(-t)$, the coefficients $b_{k}$ shown in~(\ref{eq:b}) are identified from~(\ref{eq:EIi3}) as
\begin{eqnarray}
&&-\sum_{i=0}^{2m-2}\sum_{j=0}^{i}c_{ij}\sum_{k=0}^{i+n-m}\binom{i+n-m}{k}(-t)^{i+n-m-k}k! \nonumber \\
&=&-\sum_{i=0}^{2m-2}\sum_{j=0}^{i}(i+n-m)!~c_{ij}\sum_{k=0}^{i+n-m}\frac{(-t)^k}{k!} \\
&=&-\sum_{k=0}^{n+m-2}\frac{(-t)^{k}}{k!}\sum_{j=0}^{2m-2}\sum_{i=k-n+m}^{2m-2}(i+n-m)!~c_{ij}, \label{eq:bi}
\end{eqnarray}
where the last equality follows by changing orders of the summations. Due to the presence of the factor $1/(m-1-j)!$ in $c_{ij}$, the upper bound of the sum over $j$ in~(\ref{eq:bi}) becomes $m-1$. Further simplification is possible by splitting the sum over $k$ into two sums over $k=0,\dots,n-m$ and $k=n-m+1,\dots,n+m-2$, where the sum over $i$ in the first $k$-sum now starts from $j$ due to the term $1/(i-j)!$ in $c_{ij}$. These lead~(\ref{eq:bi}) to
\begin{eqnarray}
&&-\sum_{k=0}^{n-m}\frac{(-t)^{k}}{k!}\sum_{j=0}^{m-1}\sum_{i=j}^{2m-2}(i+n-m)!~c_{ij}- \nonumber \\
&&\sum_{k=n-m+1}^{n+m-2}\frac{(-t)^{k}}{k!}\sum_{j=0}^{m-1}\sum_{i=k-n+m}^{2m-2}(i+n-m)!~c_{ij} \nonumber \\
&=&-\sum_{k=0}^{n-m}\frac{(-t)^{k}m}{k!}-\sum_{k=n-m+1}^{n+m-2}\frac{(-t)^{k}}{k!} \times \nonumber \\
&&\sum_{j=0}^{m-1}\sum_{i=k-n+m}^{2m-2}(i+n-m)!~c_{ij}, \label{eq:bi2}
\end{eqnarray}
where we have used~(\ref{eq:lnt}). From~(\ref{eq:bi2}), the coefficients of $t^{k}$ can be read off as shown in~(\ref{eq:b}). To identify the coefficients $a_{k}$, we repeatedly change the order of summations and apply the Chu-Vandermonde identity~(\ref{eq:CV}) to the corresponding term in~(\ref{eq:EIi3}) as
\begin{eqnarray}
&&\e^{t}\sum_{i=0}^{2m-2}\sum_{j=0}^{i}c_{ij}\sum_{k=0}^{i+n-m}\binom{i+n-m}{k}\times \nonumber \\
&&(-t)^{i+n-m-k}k!\sum_{s=1}^{k}\frac{\Gamma\left(s,t\right)}{s!} \nonumber \\
&=&\sum_{i=0}^{2m-2}\sum_{j=0}^{i}(i+n-m)!~c_{ij}\sum_{k=0}^{i+n-m-1}(-t)^{k}\times \nonumber \\
&&\sum_{s=1}^{i+n-m-k}\frac{(i+n-m-k-s)!}{(i+n-m+1-s)!} \nonumber \\
&=&\sum_{i=0}^{2m-2}\sum_{j=0}^{i}c_{ij}\sum_{k=0}^{i+n-m-1}\frac{(-t)^{k}}{k!}\times \nonumber \\
&&\frac{(i+n-m)!-k!(i+n-m-k)!}{k} \nonumber \\
&=&\sum_{k=0}^{n+m-3}\frac{(-t)^{k}}{k!}\sum_{j=0}^{m-1}\sum_{i=k-n+m+1}^{2m-2}c_{ij}\times \nonumber \\
&&\frac{(i+n-m)!-k!(i+n-m-k)!}{k}. \label{eq:ai}
\end{eqnarray}
From the above equation, the coefficients $a_{k}$ for $k\neq0$ can be read off as shown in~(\ref{eq:a}). For $k=0$, the equation~(\ref{eq:ai}) is indeterminant, which is resolved by using l$'$H\^{o}pital's rule as
\begin{eqnarray}
\mkern-30mu&&\sum_{j=0}^{m-1}\sum_{i=-n+m+1}^{2m-2}c_{ij}\lim_{\epsilon\to0}\frac{(i+n-m)!-\epsilon!(i+n-m-\epsilon)!}{\epsilon} \nonumber \\
\mkern-30mu&&=\sum_{j=0}^{m-1}\sum_{i=j}^{2m-2}\!\!c_{ij}(i+n-m)!\big(\psi_{0}\left(i+n-m+1\right)-\psi_{0}\left(1\right)\!\big) \nonumber \\
\mkern-30mu&&=\sum_{j=0}^{m-1}\sum_{i=j}^{2m-2}(i+n-m)!~c_{ij}\sum_{s=1}^{i+n-m}\frac{1}{s}, \label{eq:ai0}
\end{eqnarray}
where, in addition to adjust the summation bound according to $c_{ij}$, we have also utilized the definition of digamma function
\begin{equation}
\psi_{0}(x)=\frac{\dd\ln\Gamma(x)}{\dd x}=\frac{\Gamma'(x)}{\Gamma(x)}
\end{equation}
and its finite sum form for positive arguments (see, e.g.,~\cite{Wei2017})
\begin{equation}
\psi_{0}(l)=-\gamma+\sum_{k=1}^{l-1}\frac{1}{k}
\end{equation}
with $\gamma\approx0.5772$ being Euler's constant. Indeed, (\ref{eq:ai0}) is the coefficient $a_{0}$ as claimed in~(\ref{eq:a}). This completes the proof of Proposition~\ref{p:1}.

\section{Conclusion and Future Work}\label{sec:con}
We have revisited the problem of finding a closed-form formula for the ergodic mutual information of MIMO Rayleigh channels initially studied by Telatar twenty years ago. Interestingly, our motivation came from the recent progress of a similar problem in quantum information theory. We have shown that the seemingly complicated end-result by Telatar in fact admits a simple structure. The main result of this work lies on explicitly determining this simple structure, which has been summarized in Proposition~\ref{p:1}.

We finish this paper by mentioning some future research directions based on the derived results. Firstly, as a direct application we could analyze the ergodic mutual information for models with SNR variation, where the effective SNR is modeled as a random variable. The analysis is possible as the revealed dependence on SNR $1/t$ in~(\ref{eq:main}) is explicit, where, conditioned on any $t$, operations such as integration and differentiation can be easily performed. Besides modeling of physical channels, the need to take into account SNR variation comes from transceiver schemes such as modulation, coding, or channel estimation.

The considered ergodic mutual information corresponds to the first moment, whereas higher moments of mutual information are needed to describe the outage probability relevant to slow or block fading. In particular, finding the exact second moment (fluctuation) remains a long-standing open problem. Our preliminary study indicates that the higher moments also admit a similar finite polynomial structure as~(\ref{eq:main}), where the difference is that the coefficients including $a_{k}$, $b_{k}$ and the common factor $\e^{t}~\text{\normalfont Ei}(-t)$ become more complicated.

Last but not least, the proposed analytical framework can be also utilized to calculating the moments of mutual information of other MIMO channel models. For example, we expect that the Jacobi MIMO channel relevant to MIMO optical communications can be similarly studied, where, instead of the Laguerre polynomials, the Jacobi polynomials will be utilized.

\IEEEtriggeratref{3}

\appendix
\section*{Proof of Lemma~\ref{l:1}}
Denote the integral in~(\ref{eq:L1}) by $g(k)$, integration by parts gives
\begin{eqnarray}
g(k-1)&=&\frac{1}{k}\int_{t}^{\infty}\left(x^{k}\right)'\e^{-x}\ln x\dd x \\
&=&-\frac{t^{k}\e^{-t}\ln t}{k}-\frac{1}{k}\int_{t}^{\infty}\left(\e^{-x}\ln x\right)'x^{k}\dd x\nonumber \\
&=&-\frac{t^{k}\e^{-t}\ln t}{k}-\frac{1}{k}\Gamma(k,t)+\frac{1}{k}g(k), \label{eq:gRe}
\end{eqnarray}
where $f'\left(x\right)=\dd f(x)/\dd x$ and we have used the integral form of incomplete Gamma function
\begin{equation}\label{eq:iGi}
\Gamma(k,t)=\int_{t}^{\infty}x^{k-1}\e^{-x}\dd x.
\end{equation}
Equation~(\ref{eq:gRe}) implies a recurrence relation of $g(k)$,
\begin{equation}
g(k)=kg(k-1)+\Gamma(k,t)+t^{k}\e^{-t}\ln t.
\end{equation}
By iterating $k-1$ times the above relation, we arrive at
\begin{eqnarray}
g(k)\!\!\!\!\!&=\!\!\!\!\!&k!g(0)+\ln t~e^{-t}\sum_{s=0}^{k-1}\frac{k!}{(k-s)!}t^{k-s}+k!\sum_{s=1}^{k}\frac{\Gamma\left(s,t\right)}{s!} \nonumber\\
&&\mkern-60mu = k!g(0)-k!e^{-t}\ln t+\Gamma(k+1,t)\ln t+k!\sum_{s=1}^{k}\frac{\Gamma\left(s,t\right)}{s!}, \label{eq:gk}
\end{eqnarray}
where the second equality is obtained by~(\ref{eq:iG}). Using integration by parts, the integral $g(0)$ is calculated as
\begin{eqnarray}
g(0)&=&\int_{t}^{\infty}\e^{-x}\ln x \dd x \\
&=&\e^{-t}\ln t+\int_{t}^{\infty}\frac{\e^{-x}}{x}\dd x \\
&=&\e^{-t}\ln t+\Gamma(0,t), \label{eq:g0}
\end{eqnarray}
where we have used the definition~(\ref{eq:iGi}) for $k=0$. Inserting~(\ref{eq:g0}) into~(\ref{eq:gk}), we arrive at the claimed result~(\ref{eq:L1}).

\end{document}